\documentstyle[epsf]{l-aa}
\begin{document}

\thesaurus{08 % A&A Section 8: Stars
            (08.14.1 % neutron
		) 
           13 % A&A Section 13: Gamma rays
            (13.07.1 % bursts
		)
          }    
\title{
Formation of a Gravitationally Bound Object\\
after Binary Neutron Star Merging and GRB phenomena}

\author{G.V.~Lipunova\inst{1} \and V.M.~Lipunov\inst{2}
       }
\offprints{G.V.~Lipunova}

\institute{Sternberg Astronomical Institute, Moscow State University,
        Universitetskii pr. 13,
	Moscow, 119899 Russia\\
	email galja@sai.msu.su
 	\and
	Sternberg Astronomical Institute, Moscow State University 
        and Milano University\\
	email lipunov@sai.msu.su}
\date{Received ~~~~~~~~~~~~~~~~~~~ /Accepted}
\maketitle
\markboth{G.Lipunova, V.Lipunov.  Neutron Star Merging and GRB phenomena}{ ...}

\begin{abstract}

The stages that follow  the merging of two neutron stars are discussed.
It is shown that if a rapidly rotating gravitationally bound object is
formed after the merging (a spinar or a massive neutron star), then the 
characteristic time of its evolution is determined by a fundamental value

$$ t_{spin} = \kappa {{m_p^2 e^2 \hbar^{1/2}}\over m_e^4 c^{5/2} G^{1/2}}
   \approx 7\cdot 10^5 s \cdot \kappa~,  $$
where the dimensionless value $\kappa$
%=100\div 1000$  
depends on the exact equation of state of nuclear matter. 
The hypothesis is discussed as to whether the residual
optical emission of the gamma-ray bursts 
is pulsar-like and its evolution driven by magnetodipole energy losses.
It is shown that binary neutron  star mergings
can be accompanied by two gravitational wave  burst separated either
by the time of spinar's collapse $t_{spin}$ or neutron star 
cooling time ($\sim$ 10 s),
depending on the masses of neutron stars.


\end{abstract}

\section{Introduction}
The detection of optical and X-ray emission after the \break gamma ray
bursts GRB~970228, GRB~970508 
(Groot et al. 1997a; Groot et al. 1997b; Metzger et al. 1997b ;
Costa et al. 1997b; Sahu et al. 1997; Bond 1997; Galama et al. 1997;
Djorgovski et al. 1997a; Metzger et al. 1997c; Schaefer et al. 1997;
Djorgovski et al. 1997b; Djorgovski et al. 1997c; Groot et al. 1997;
Donahue et al., 1997) 
may be interpreted in terms of the formation of a 
transient rapidly rotating gravitationally bound object ---
a heavy neutron star (NS) or spinar --- an object with the equilibrium
maintained either by
the fast rotation (``cool'' spinar, $CSP$) or by 
both  rotation and pressure
(``hot'' spinar, $HSP$).

  Let us assume that two neutron stars with masses $ M_1 $ and $ M_2 $
are merging. The following state of the after-merging object 
is determined by the ratio
of the resulting total mass and the Oppenheimer--Volkoff limit.
Two different scenarios may be envisaged as follows:

$$     M_1 + M_2 \geq M_{OV}          \eqno (\hbox{A}) $$
$$     M_1 + M_2 < M_{OV}          \eqno (\hbox{B}) $$

Here and below we interpret the Oppenheimer--Volkoff limit not as the
standard value derived for the cold equation of state of baryonic matter
for a non--rotating neutron star, but as a modified one.
In the general case the Oppenheimer--Volkoff limit is a function of
the angular spin velocity of the object, its entropy, and the specific
equation of state: $M_{OV}=M_{OV}(\omega, S, EqSt) $.

Each neutron star can have a mass lying between the limits:

$$             M_{min}   <  M_1, M_2  <  M_{OV}        $$

  The value of $M_{min} \sim  0.2 M_\odot$ was derived by Landau (1938). 
In a standard
modern scenario, it is commonly suggested that $M_{min} \sim 1.2 M_{\odot} $.
Thus we can expect the different evolutionary tracks depending on
the specific masses of NS.

\section{Mergingology}
\subsection{ Case (A)}

In this case, we can expect that after the merging a black hole
results from a direct collapse during the time $\sim 10^{-5} s$ and
that the most energy is emitted in the gravitational wave burst.
This scenario is discussed more frequently in the literature, and 
GRB phenomenon can be related with the relativistic particle ejection
in the form of a Fireball (Rees \& Meszaros, 1992) or a beam of protons
(Shaviv \& Dar, 1996).
In addition, a certain fraction of radiated energy can be related with the
pulsar mechanism (Lipunov \& Panchenko 1996; Lipunova 1997).
No gravitationally bound object can be formed in this case outside
the horizon.  We can present these stages by the following way:

$$ NS + NS \rightarrow BH + GWB + GRB +\nu B $$
($GWB$ - gravitational waves burst; $\nu B$ - neutrino burst).

From our point of view the more interesting scenario is the Case (B).

\subsection{ Case (B): $M_1+M_2 < M_{OV}$ }

This variant can be realized if either two merging neutron stars have
small masses or the Oppenheimer-Volkoff limit is very large.

Is it possible that the Oppenheimer--Volkoff limit exceeds
$3M_{\odot}$? First, it is known (Friedman \& Ipser, 1987)
that the fast rotation (which is naturally expected after the merging)
increases the Oppenheimer--Volkoff limit to the value
$\sim 3 M_{\odot}$ for hard equations of state.
Second, the object formed is not degenerate due to its high
temperature and the equilibrium is maintained both by fast rotation
and entropy (``hot'' spinar). And last, Oppenheimer-Volkoff
limit can be high because of relativistic behavior of nuclear
forces.

Thus we can present these three sub-scenarios as follows:

$$ NS + NS \rightarrow HSP + GWB + GRB +\nu B $$
$$ NS + NS \rightarrow CSP + GWB + GRB +\nu B $$
$$ NS + NS \rightarrow NS + GWB + GRB +\nu B $$

Let us consider the case  of the HSP. Its lifetime is completely
determined by the cooling time which, according to different calculations,
is of the order of $\sim 10$~s (Shapiro \& Teukolsky, 1983). Then, in the time
interval $t_{cool}$, the collapse accompanied by the GWB, neutrino emission,
and possible weak photon emission can be expected:

$$ HSP \rightarrow BH + GWB + \nu B + \gamma $$

It seems very attractive to identify this cooling time with the
mean characteristic gamma-ray burst duration $\sim 1\div 10$~s!

Second, the  most interesting sub-scenario is when the centrifugal forces
make the main contribution to the equilibrium  (``cool'' spinar).
In this case the lifetime of the spinar is completely defined
by the characteristic time of the angular momentum loss $t_{spin}$ and
evolutionary track looks like
$$ CSP \rightarrow BH + GWB + \gamma + e^+ + e^- + \nu $$

Finally, there is a case of a high Oppenheimer-Volkoff limit for the
cool non--rotating object.

$$  M_1 + M_2 < M_{OV} \hbox{~~~~~~~~~~~~~ always~! }    $$

This variant leads to the formation of a very powerful pulsar (maybe 
without pulsation) with the maximum spin rotation.

$$   NS + NS \rightarrow  PSR       $$

The characteristic time $t_{spin}$ of its evolution is governed by the
momentum loss rate.

\section{The rate of the angular momentum losses}

In both cases of a cool spinar (Lipunova 1997) and of a fast--rotating $NS$,
the specific time of their evolution is determined by the rate of magnetodipole
energy loss

$$ {dI\omega\over dt} = -{2\over 3} {{\mu^2 \omega^3}\over c^3}\, , $$

and

$$ t_{spin} = {\omega\over 2\dot\omega} =
            {2\over 5} {M c^3\over B_o^2  R_o^4 \omega^2}\, .        $$
We assume:
\par
the inertia moment \hskip 2cm $  I = {2\over 5} M R^2\, ,           $
\par
the mass           \hskip 3cm $ M = M_1 +M_2\, ,                $
\par
the magneto--dipole moment~~ $\mu = B_o R_o^3/ 2\, . $
\par
\noindent
The angular spin velocity of the post-merging object must
be close to the limit:

$$        \omega = (GM/R_o^3)^{1/2}\, .       $$

Then we obtain:

$$       t_{spin} \approx {6\over 5}{c^3\over B_o^2 G R}
      \approx 2\cdot10^5 \left({B\over B_{cr}}\right)^{-2}
                    \left({R\over 10^6 cm}\right)^{-1} \hbox{s}~ . $$

Thus, this duration is determined mainly by the magnetic field.
If we assume that a gravitationally bound object magnetic field is equal
 to the
critical value close to the Schwinger limit:

$$ \hbar {{e B_{cr}}\over m_e c} = m_ec^2, \quad
          B_{cr}\approx 4.3\cdot 10^{13}~ \hbox{G~.}   $$

Expressing the radius and the mass of the NS
in terms of fundamental constants we obtain the
fundamental value for the lifetime of such an object:

$$ T= {{m_p^2 e^2 \hbar^{1/2}}\over m_e^4 c^{5/2} G^{1/2}}
     \approx 7.6 \cdot 10^5 \hbox{s}~.  $$

Taking into account the real mass of NSs and specific equation of state,
this time can be modified as

$$         t_{spin} = T \cdot \kappa~, $$
where $\kappa$ depends on the exact equation of state of nuclear matter.
%\approx 100\div 1000 $. 
This duration accords with the specific
fundamental value of luminosity.

\begin{figure}[t]
 \epsfxsize=0.45\textwidth
 \begin{center}
  \mbox{\epsfbox{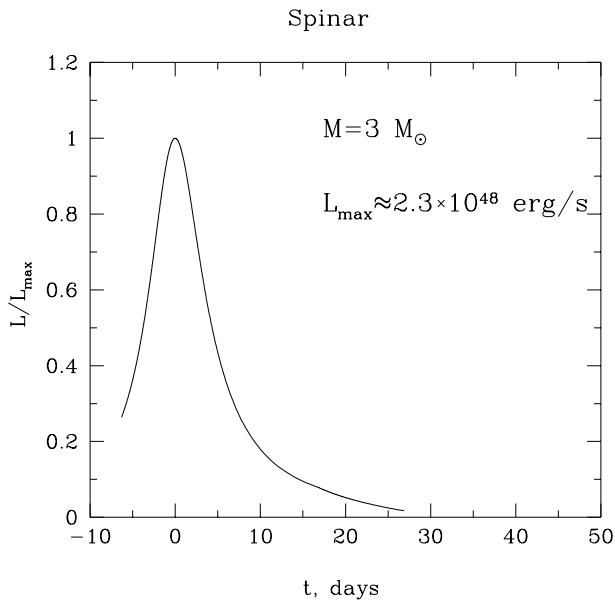}}
  \caption{The spinar luminosity evolution.
The magnetic field $B=4.3\cdot 10^{13}$~Gs at $R=20$~km.}
 \end{center}
\label{Fig1}
\end{figure}

\section{GRB light curve}

\begin{figure}[t]
\vspace{4mm}
 \epsfxsize=0.45\textwidth
 \begin{center}
  \mbox{\epsfbox{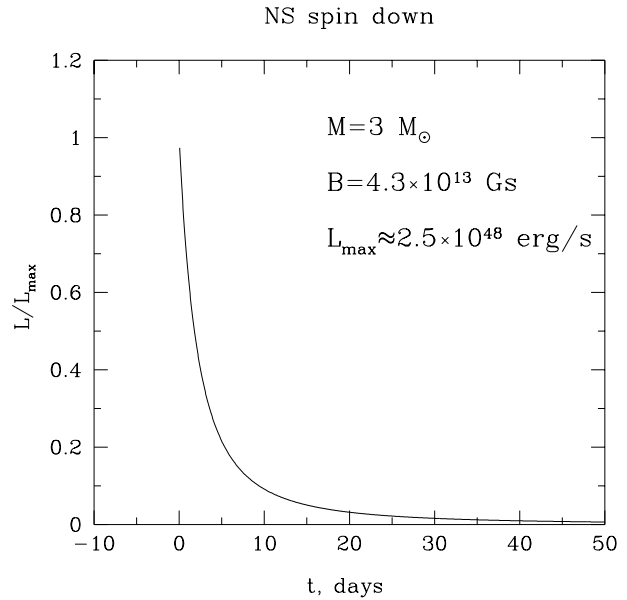}}
  \caption{The neutron star spindown rate and  luminosity evolution. 
$R=30$~km, $\nu_{max} = 660$~Hz. }
 \end{center}
\label{Fig2}
\end{figure}

Here we present the alternative model to the now frequently discussed
model of GRB --- the radiation of a Fireball (see Meszaros \& Rees 1997). 
We admit that the models of fast--rotating pulsar or spinar with
extremely high magnetic fields do not wholly substitute the model 
of Fireball (especially, concerning the gamma burst itself) 
but accompany the process of radiation and, possibly,
at later stages of GRB afterglow, dominate in a GRB spectrum.
Note, that these mechanisms can supply the emission in a wide
range of wavelengths, as radiopulsar studies confirm.

We suggest that part of the observed optical and X-ray
afterglow of a GRB can relate to the pulsar mechanism.
As an emission from Fireball decreases to the undetectable level, 
the pulsar mechanism can become the main contribution to the 
afterglow.
 
We can construct the luminosity evolution for a cool spinar collapse
(Lipunova, 1997) and for a pulsar spin down.

Supposing that the optical emission is produced by the
pulsar mechanism acting with the critical magnetic field, one can derive:
$$ L\approx 2\cdot 10^{45} \mbox{erg}/\mbox{s} 
\left({B\over B_{cr}}\right)^2
    R_6^6 P_{1.5}^{-4}~ K(t , \nu)~\times$$
$$\times ~\left({t\over 3\cdot 10^7~\mbox{s}}
\left({B\over B_{cr}}\right)^2 R_6^4
P_{1.5}^{-2} M_3^{-1}+1 \right)^{-2}~,$$
where
$R_6=\left({R/ 10^6 cm}\right)$,
$P_{1.5}=\left({P/1.5 ms}\right)$,\break
$M_3=\left({M/ 3 M_{\odot}}\right)$.
The  coefficient $ K(t , \nu)$ is the ratio of optical radiation to
the total energy loss by a pulsar.
Of course, it is hard to expect the ratio of optical radiation  
to the total rotational energy loss to be constant, as evidenced
by radiopulsar studies. As it is, the real power of time dependence
can vary from $-2$.

Fig.~1 shows the characteristic times of luminosity decreasing to be
in a rather good correlation with the observed 
ones  (see Groot et al. 1997a; Groot et al. 1997b; Metzger et al. 1997b
Costa et al. 1997b; Sahu et al. 1997;
Bond 1997; Galama et al. 1997;
Djorgovski et al. 1997a; Metzger et al. 1997c; Schaefer et al. 1997;
Djorgovski et al. 1997b; Djorgovski et al. 1997c; Groot et al. 1997;
Donahue et al., 1997).
The model of a neutron star spin down is calculated for the initial 
angular velocity
$\nu = 660$~Hz, which  corresponds to the minimum spin period observed
in  millisecond pulsars. 

The lack of optical counterparts to other GRBs may be explained
by another relation between the total mass of the system before merging and
the Oppenheimer-Volkoff limit and, as a result, by another
scenario of neutron star coalescence. 

The authors acknowledge Dr K.A.Postnov and All-Moscow Seminar of 
Astrophysics (OSA, http://xray.sai.\break{msu.su/sciwork/eseminars.html)}
for valuable discussions.

\bigskip
\bigskip

The work is supported by Cariplo Foundation for Scientific                            
Research and by Russian Fund for Basic Research through Grant No 95-02-06053.


\begin{thebibliography}{}
\bibitem{}Bond H.E., 1997, IAUC 6654
\bibitem{}Costa E., Feroci M., Frontera F. et al., 1997a, IAUC 6572
\bibitem{}Costa E., Feroci M., Piro L. et al., 1997b, IAUC 6576
\bibitem{}Galama T.J., Groot P.J., 1997, IAUC 6655
\bibitem{}Groot P.J., Galama T.J., van Paradijs J. et al., 1997a, IAUC 6584
\bibitem{}Groot P.J., Galama T.J., van Paradijs J. et al., 1997b, IAUC 6588
\bibitem{}Groot P.J., Galama T.J., van Paradijs J. et al., 1997c, IAUC 6660
\bibitem{}Dar A. \& Shaviv N., 1996, astro-ph/9607160
\bibitem{}Djorgovski S.G., Metzger M.R., Odewahn R.R. et al., 1997a, IAUC
6655
\bibitem{}Djorgovski S.G., Odewahn R.R., Gal R.R. et al., 1997b, IAUC 6658
\bibitem{}Djorgovski S.G., Metzger M.R., Kulkarni S.R. et al., 1997c, IAUC 6660
\bibitem{}Donahue M., Sahu K.C., Livio M. et al., 1997, IAUC 6666
\bibitem{}Friedman J \& Ipser R., 1987, ApJ 314, 594
\bibitem{}Lipunova G.V., 1997, Astronomy Letters, 23, 104
\bibitem{}Lipunov V.M. and Panchenko I.E., 1996, Astronomy and Astrophysics 312, 937
\bibitem{}Lipunov V.M., Postnov K.A., Prokhorov M.E.,
Panchenko I.E., Jorgensen H., 1995, ApJ 454, 593
\bibitem{}Lipunov V.M., Postnov K.A., Prokhorov M.E.,1997, astro-ph/9703181
\bibitem{}Mallozzi E. et al., 1996, ApJ 471, 636
\bibitem{}Meszaros P. \& Rees M., 1997, ApJ 476, 232
\bibitem{}Metzger M.R., Kulkarni S.R., Djorgovski S.G.
et al.,1997a, IAUC 6582
\bibitem{}Metzger M.R., Kulkarni S.R., Djorgovski S.G. et al., 1997b, IAUC 6588
\bibitem{}Metzger M.R., Djorgovski S.G., Steidel C.C. et al.,1997c, IAUC
6655 
\bibitem{}Piro L., Costa E., Feroci M. et al., 1996, IAUC 6467
\bibitem{}Rees M. \& Meszaros P., 1992, MNRAS 258,.41
\bibitem{}Sahu K., Livio M., Petro L., Macchetto F.D., 1997,
IAUC 6606
\bibitem{}Schaefer B., Schaefer M., 1997, IAUC 6658 
\bibitem{}Shapiro S. \& Teukolsky S., 1983, Black Holes, White Dwarfs
and Neutron Stars, A Wiley-Interscience Publication, New-York
\end{thebibliography}
\end{document}